\documentclass[aps]{revtex4}
\usepackage{epsfig}

\begin{document}

\begin{flushright}
Edinburgh 2016/12
\end{flushright}

%
%
%
%
%

\bigskip


\title{Space dust collisions as a planetary escape mechanism}

\author{Arjun Berera}
\email{ab@ph.ed.ac.uk}

\affiliation{School of Physics and Astronomy,
University of Edinburgh, Edinburgh EH9 3FD, United Kingdom}





\begin{abstract}
It is observed that hypervelocity space dust, which is
continuously bombarding the Earth,
creates immense momentum flows in the atmosphere.
Some of this fast space dust inevitably will
interact with the atmospheric system, transferring energy and
moving particles around, with various possible consequences.
This paper examines, with
supporting estimates, the possibility that through collisions,
the Earth-grazing component of space dust can
facilitate planetary escape of atmospheric particles,
whether they be the atoms and molecules forming
the atmosphere or bigger sized particles.
As one interesting outcome, floating in
the Earth's atmosphere are a variety of particles containing
the telltale signs of Earth's organic story, including microbial
life and life essential molecules.  This paper will assess the
ability for this
space dust collision mechanism to propel
some of these biological constituents into space.
\end{abstract}

\maketitle

\bigskip

In press {\it Astrobiology}, 2017




%
%

\section{Introduction}

A huge amount of space dust enters
the Earth, on the scale of $~\approx 10^5$ kilograms per day, 
that is composed of
dust particles of varying masses from $10^{-18}$ to $1$ gram
and enters the Earth's atmosphere at very high speeds
$~\approx 10-70 {\rm km/s}$ \citet{kw86,lb93,flynn2002,csfbkch,plane}.
This hypervelocity space dust forms immense and
sustained momentum flows in the
atmosphere.
For particles that form the thermosphere or above or reach there
from the ground, if they
collide with this
space dust, they can be displaced, altered in form or carried off
by incoming space dust.  
This may have consequences for weather and wind, but
most intriguing and the focus of this paper, is the
possibility that
such collisions can give
particles in the atmosphere
the necessary escape velocity
and upward trajectory to escape Earth's gravity.
Two types of particles that will
be considered are either light elements/molecules
that form Earth's atmosphere or bigger particles
capable of harboring life or life essential molecules.
The former possibility implies an
exchange mechanism of atmospheric constituents amongst widely separated
planetary bodies. The latter, and perhaps most interesting possibility,
addresses basic questions about the
origin of life, with similarities to the
Classical Panspermia mechanism \citet{arrhenius1908}, except space dust
rather than radiation scrapes up life in the upper atmosphere.
One should approach the
application of this space dust collision mechanism to
panspermia cautiously, since there are several complicating factors
and they will be considered in this paper. 
However the prospects are intriquing and
so worth exploring.
Earth harbors the greatest, perhaps only, concentration of life and
its biologically produced molecules within this
local region of our Galaxy.
The idea that microbial life and its
constituents are exchanged between planets
would gain further scientific footing
by establishing that natural processes on Earth 
send out its biological constituents
into the Solar System.

Realizing gravitational escape for small
particles  presents a few difficulties.  First it 
requires upward forces that can 
accelerate these particles up to escape velocity level.  On the one
hand, if this is done at too low an altitude, the stratosphere or below,
the atmospheric density is so high that drag forces will
rapidly slow fast moving particles.  Moreover these particles
will also undergo immense heating to the point of even evaporating.
For these reasons, even though wind, lightning, volcanoes etc... all would
be capable of imparting huge forces at these lower altitudes,
they would not be able, even in principle, to
accelerate particles intact up to escape velocity.
On the other hand, at very high altitudes, at the
upper part of the mesosphere and into the thermosphere, 
particles moving at
escape velocity levels would not suffer such great drag and heating
effects and so could escape the Earth's gravity and cruise into
outer space.  As such, only in the higher atmosphere
would it even be possible that the atoms and molecules
found there could be propelled into
space by space dust collisions.  As for larger particles
capable of harboring biological constituents,
the most likely scenario for thrusting them into space
would require a double stage approach, whereby they are first hurled 
into the lower thermosphere region or higher by some
mechanism and then
given an even stronger kick by fast space dust collision,
which eventually leads to escape
velocity and an exit from the Earth's gravity.

\subsection{Drag force and surface temperature}

Consider
the forces that act on a particle of mass $m$ at an altitude
$z$ above sea level.  First there is the downward gravitational force,
which for altitudes within a few hundred kilometers of
sea level, can be approximated as just a constant
$m g$ with $g \approx 9.8 {\rm m/s^2}$. Second is the drag force
in the direction opposite the motion of the particle,
\begin{equation}
\frac{d v}{dt} = -\frac{3 \Gamma \rho_a}{4 \rho r} v^2 \;,
\label{dragv}
\end{equation}
where $v$ is the speed of the particle, 
$\rho_a$ is the atmospheric mass density,
$\rho$ is the particle mass density, $r$ is the particle radius,
and $\Gamma \approx 1$ is the atmospheric drag coefficient.
The atmospheric density \citet{hkl52,nrlmsise00,brekke13}
as function of altitude $z$ is
$\rho_a(z) \approx (1.2 \times 10^3 {\rm g}/{\rm m}^{3}) \exp\left(-{z}/{7.04}\right)$,
which is valid to within approximately a factor $2-3$ and up
to an altitude around $150 {\rm km}$.
Third a particle at speed $v$ will get heated 
from collisions with the molecules in the air as
\begin{equation}
\alpha \frac{1}{2} \rho_a 4 \pi r^2 v^3 = 4 \pi r^2 \sigma T^4 \;,
\label{tempeq}
\end{equation}
where the LHS is the rate of energy gained by the particle from the
kinetic energy of the molecules in the air and the RHS is the blackbody rate
at which the particle radiates the energy, leading to a surface
temperature $T$.  Here $\sigma$ is the Stefan-Boltzmann constant and
$\alpha$ is the fraction of the total kinetic energy of the
air molecules that stick to the moving particle, which
we will set to $\alpha = 1$.  There can also be a heat capacity term
on the RHS of Eq. (\ref{tempeq}), but for small particles of
micron size or a couple of orders of magnitude bigger,
this term can be ignored \citet{whipple50}

\section{Estimates}

Let us first make some estimates on the speed that
small particles, whether they be the space dust or
particles sitting in the atmosphere, are able to achieve
in the atmosphere, accounting for drag and gravitational
forces and surface temperature levels.
For atmospheric elements and molecules,
the scenario proposed in this paper is that
as the space dust moves through the atmosphere, some
of the atoms and molecules that it is made of get
stuck to the space
dust. Since these atoms and molecules will be
of negligible mass compared to the space dust, after collision the space
dust will simply maintain the same speed it had.  For bigger
particles capable of harboring microbial life, the collision
dynamics can range from a simple elastic collision,
to the small particle sticking to the space dust particle,
to fragmentation/vaporization of one or both particles in the collision. 
Note however this paper focuses on interplanetary space
dust which has a velocity range when coming to Earth
of $~\approx 10-70 {\rm km/s}$.  Collisions between
small grains in this velocity range have been shown in
\citet{bd95} to not completely destroy them.
Some percent of the colliding grains do undergo fragmentation/vaporization,
but some portion also remains intact. At the same time,
since the velocity range for the fast space dust is still
some factors higher than escape velocity,
provided the incoming space dust particle is the same or bigger
size than the small particle in the atmosphere it collides with,
both the case of elastic collision and where the atmospheric particle
sticks to the incoming space dust particle will result in 
a final velocity still around the same as the incoming velocity of
the space dust.  Although some fraction of the two colliding
particles may fragment or vaporize \citet{bd95}, given the 
huge energy being transferred in this process, it still is
plausible that portions of the two colliding
particles ultimately emerge with high velocity,
capable of escaping the Earth's gravity.

\subsection{Speed and altitude}

To make some estimates, for the space dust, it has
typical density $\rho_s \approx 2 \times 10^6 {\rm g}/{\rm m}^{3}$.
The small atmospheric particles struck by this space dust may
contain life related quantities,such as bacteria, viruses, fungi,
life related molecules like DNA, RNA, etc...
For example a typical bacteria has a mass
around $\sim 10^{-12} - 10^{-13} {\rm g}$ and
and  a length around $r \approx 10^{-6} {\rm m}$.
Other biological constituents will have approximately
similar density.
The small atmospheric particles will be approximated
as spherical and have
a density $\rho_p$ the same as the space dust.

\begin{figure}[h]
 \centering
%
\includegraphics[width=31pc]{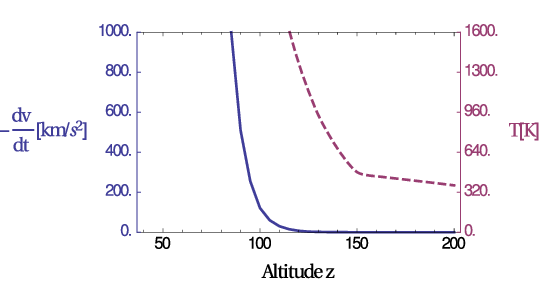}
%
\caption{Deceleration due to atmospheric drag forces
and surface temperature as function of
altitude for a particle of radius $r = 10^{-6} m$ moving through the atmosphere at speed of $20 {\rm km/s}$}
\label{fig1}
\end{figure}

Focusing first at higher altitudes,
the speeds of interest here are near escape velocity,
so in the ${\rm km/s}$ range,
or within an order of magnitude around this.
As the interest is in vertical motion, atmospheric densities
experienced by these fast moving particles at these speeds
will change substantially
within timescales of around a second.
The effect of the gravitational acceleration is in
the ${\rm m/s^2}$ range and so can be ignored, since it
will be a small effect on particles of ${\rm km/s}$ speeds. 
For a particle of radius
$r = 10^{-6} {\rm m}$ and speed $20 {\rm km/s}$,
the results for the deceleration from Eq. (\ref{dragv}) and surface
temperature from Eq. (\ref{tempeq}) are shown in 
Figure~\ref{fig1}.  In particular
the deceleration from atmospheric drag at
altitudes $110,130,$ and $150 {\rm km}$
from Eq. (\ref{dragv}) respectively is
around (in ${\rm km/s^2}$) $-30,-2$, and $-0.1$,
as shown in Figure~\ref{fig1}.
These expressions scale quadratically with speed, so for
every factor two less in speed, these expression decrease by a factor
four. Moreover they scale inverse with the radius, so for every order
of magnitude increase in $r$ these expressions decrease by
an order of magnitude.  Estimating also the
surface temperature from Eq. (\ref{tempeq})
at the same speed of $20 {\rm km/s}$
for the same
altitudes $110,130$, and $150 {\rm km}$, gives respectively
$~\approx 2000,1000, 500$ K as shown in Figure~\ref{fig1}.  
These expressions scale to the $3/4$th power 
with speed
and are independent of the particle radius.
So for example at $150 {\rm km}$ altitude, if the speed is
around a factor two smaller, so just around escape
velocity $11.2 {\rm km/s}$, 
the surface temperatures will be within
the range where biological life can be sustained.
From these simplified expressions, it suggests the minimum
altitude of around $150 {\rm km}$ above which 
a single collision to escape velocity can allow
a particle to cruise free of the Earth's gravity, unhindered
by atmospheric drag and heating effects.

At much lower altitudes, our estimates show it is very difficult
to accelerate particles to any substantial speed.  For example
at an altitude of $20 {\rm km}$, where micron and larger
sized particles
are known to be found in the atmosphere with some containing microbial life, 
the drag force and surface temperature remain
adequately small only for
speeds at most $~\stackrel{<}{\sim}~20~{\rm m/s}$
and the particle radius needs to be in the millimeter or larger range.
This speed is low enough that the gravitational force
is relevant, and will pull such particles down unless they
experience frequent upward forces to maintain this upward speed.
At around $50 {\rm km}$ altitude, drag forces and
surface temperature are small for
particles with speed  $~\stackrel{<}{\sim} 30 {\rm m/s}$
and radius larger than around $10^{-5} {\rm m}$.
At around $85 {\rm km}$, which is approximately the highest altitude
that particles from volcanic eruptions are believed to have 
reached \citet{verbeek1884,sr81,ludlam57}, 
drag forces and surface temperatures become small for speeds
$~\stackrel{<}{\sim}~0.5~{\rm km/s}$ and particle
radius larger than $10^{-6} {\rm m}$.

\subsection{Flux of hypervelocity space dust with grazing trajectories}

Next we need to consider what is the rate at which fast moving
space dust collides with particles in the atmosphere.  We are
interested in collisions that accelerate particles upward to
higher altitude and eventually escape from the Earth's gravity.
Space dust is bombarding the Earth from all directions. 
Although much of of this dust will get pulled down
by Earth's gravity and fall to the ground, given the high entry speeds of
this dust in the order of and even much bigger 
than escape velocity levels, some of
the dust that enters the Earth's atmosphere will then
graze through it.
For larger sized meteorites, such phenomenon is well known and
is visible in spectacular fireballs that streak through the
sky, sometimes accompanied also with meteor showers.  Some fraction of
the lighter space dust will also simply pass through the Earth's
atmosphere.  For a space dust particle that is grazing past the Earth,
just beyond the point where it moves exactly
parallel to the ground beneath it, this space dust 
will be moving with an increasing
upward incline relative to the ground immediately below it.  
It is beyond this point that if it collides with particles in
the atmosphere it will give them an upward force, accelerating them
to higher altitudes.  At very high altitudes around $150 {\rm km}$ and higher,
we saw from the above estimates that at escape velocity level,
drag and heating effects are not significant.  Above this
altitude, fast moving space dust will not heat up significantly and
will continue to move fast.  
Anticipating that the chances of space dust hitting small particles
in the atmosphere is a rare event, which will be verified
below, it will be
assumed that a given atmospheric particle may have at most
one or two collision
with space dust. If this is to provide adequate momentum to
the particle, the fast space dust needs to be at least the same or greater
mass than the particle it hits. In such collisions, by momentum
conservation, a considerable portion of debris after collision
will then leave with about the same speed and direction as the incoming
fast particle that initially hit it.
If such a collision by a fast moving space dust particle with
some upward velocity happened at high enough altitude, then
the struck particle, whether attaching itself to the incoming
space dust particle or scattering elastically/semi-elastically,
could be accelerated
to escape velocity level.
If that small particle contained any biological constituents,
these would be thrust out into space, free of the Earth's gravity.

Alternatively at lower altitudes,
where atmospheric drag forces and heating effects prohibit particles
from moving very fast, space dust with some upward momentum
colliding with
particles in the atmosphere would accelerate them 
to sub-escape velocity levels, but
possibly still high enough to push the particles substantially further up in
the atmosphere.  Once higher up, a second collision could then
serve to propel the particle out into space.
This possibility at lower altitudes is by no means the only
mechanism for elevating particles upward and may not even
be a significant mechanism to achieve this.  Other forces
are known to provide particles with upward velocity from
lower altitudes such as weather phenomenon, 
atmospheric electric fields during thunderstorms
\citet{pasko2002,dld},
gravito-photophoresis \citet{rohatschek96}, 
mesospheric and thermospheric 
vertical wind \citet{wg74,rb76,rscmla84,ketal09,ermr11}
and volcanic eruptions \citet{ludlam57,wshw78,tthgr09}.  
Nevertheless since the
effects of collisions with fast moving space dust is
being examined, for
completeness this process at lower altitudes is also considered.

\begin{figure}[h]
 \centering
%
\includegraphics[width=31pc]{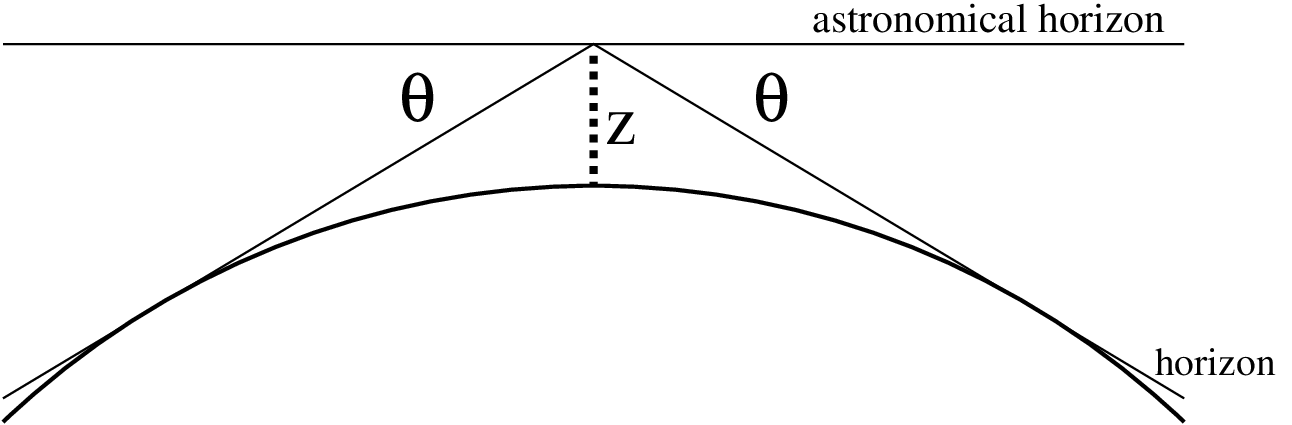}
%
\caption{Earth-grazing space dust approaches point at altitude $z$ from within
angle $\theta$ and has some upward component of momemtum}
\label{fig2}
\end{figure}

In order to estimate the chance of occurrence of space dust
collisions with particles in the atmosphere, 
it requires the flux of space
dust moving upward away from the Earth at a given point that
is a normal distance
$z$ above the ground, as shown in Figure~\ref{fig2}. 
At this zenith point consider  
the plane that is parallel to the Earth's surface 
just beneath it, also known as the plane defined by the astronomical
horizon around this point.  Space dust particles coming to this
zenith point from beneath this plane will have some upward
component of momentum.  From this zenith point, lines extending
tangent to the Earth's surface, thus to the true horizon, in all
directions define a cone.  The region between
the astronomical horizon plane and this cone section, as
shown in Figure~\ref{fig2}, is the maximum
region in which Earth-grazing space dust approaching this zenith point
will have some upward component of velocity. 
The angle between the astronomical plane and this cone is
$\theta \equiv \sqrt{2z/R_e} \ll 1$, 
where $R_e \approx 6400 {\rm km}$ ($ \gg z$) is the
radius of the Earth.
The space dust flux that can pass this zenith point from all
directions forms
a solid angle region of $2 \pi + 2 \pi \theta$.
Assuming a uniform distribution of space dust flux 
from all directions heading to this zenith
point, then the
approximate fraction of this flux
with some component of upward velocity
is $~\approx \sqrt{2z/R_e}$. 

Space dust that is not moving upward can still
impart transverse momentum on an atmospheric particle
during a collision that could have some upward component.
Also for space dust moving toward the point z within an angle $\theta$ above
the astronomical horizon plane, a forward collision of
it with an atmospheric particle will give it an Earth-grazing trajectory
which initially moves downward with respect to the ground beneath it but
eventually will move upward.  Another possibility is
when the fast space dust collides
with atmospheric particles, they could fragment from the high impact
collision.  Given that the space dust speed can be
many factors higher than the required escape velocity, even 
if the collision results in a fragmentation of the atmospheric particle,
there is adequate kinetic energy present in such collisions that
some emerging fragments could still be capable of
having speed at escape velocity level.
For the order of magnitude estimates of
interest here, all these details will not be treated.  

There has been considerable work in determining the amount and mass
distribution of space dust bombarding the Earth 
\citet{kw86,lb93,flynn2002,csfbkch,plane}.
Measurements show 
that the Earth receives roughly $10^7 - 10^8$ kilograms per 
year of space dust.  Estimates from ground and satellite
measurements also show that the flux to the Earth is
in the neighborhood of about $10^9$ grams of space dust
per year for each decade of particle mass from $10^{-9} {\rm g}$ to
$10^{-2} {\rm g}$ \citet{kw86,lb93,flynn2002,plane}.  Based on
the arguments given above, the fraction of this space dust flux
that will have some upward velocity is for example
at $z=20 {\rm km}$, $\sqrt{2z/R_e} \approx 0.08$,
whereas at $z=150 {\rm km}$ it is $\sqrt{2z/R_e} \approx 0.2$.

\subsection{Planetary escape of small particles in the higher atmosphere}

Examine a particle sitting high in the atmosphere
at altitude $150 {\rm km}$ with mass $10^{-11} {\rm g}$.  Assuming
a typical matter density of $2 {\rm g/cm^3}$ this implies
a particle radius $r_p \approx 10^{-6} {\rm m}$.  This mass and
size correspond to, for example, a collection of several bacteria or
some bacteria packed in some dirt or water.
If a fast moving space dust particle that is grazing the Earth 
at this altitude with
mass at or above that of this small particle, collides with it, the
result will be this small particle then emerges with approximately the same
speed and direction as the incoming space dust particle.
Using the measurements of space dust flux by 
\citet{kw86,lb93,flynn2002,csfbkch,plane},
space dust with radius $10^{-6}{\rm m}$, $10^{-5} {\rm m}$, 
$10^{-4} {\rm m}$, and $10^{-3} {\rm m}$
have particle flux, $f$, in ${\rm m}^{-2} {\rm s}^{-1}$ of 
order $10^{-3}$, $10^{-6}$, 
$10^{-8}$,
and $10^{-12}$ respectively. 
Based on our above estimates, at the altitude of $150 {\rm km}$,
these numbers need to be multiplied by
$0.2$ to obtain the corresponding flux, $f_u$, 
of space dust particles with some
upward velocity.
The rate at which this upward fast moving space dust will
collide with a particle in the atmosphere is $R = \sigma f_u$, where
$\sigma = \pi (r_p + r_s)^2$ is the classical hard sphere cross
section between the particle $p$ and the space dust $s$.
Calculating, we find at $150 {\rm km}$ altitude
the collision rate is dominated by space dust particles
of radius at or below $r_s \approx 10^{-4} {\rm m}$, with
$R \approx 2 \times 10^{-15} {\rm s}^{-1}$.  To put this
value for the rate in perspective, if there was one
atmospheric particle with radius at or less than $\approx 10^{-6} {\rm m}$
within each meter squared column of the atmosphere all
around the Earth at or above
this altitude, it would lead to about one such particle being accelerated
upward
to escape velocity every second.  To see this another way, if we ask
what is the smallest number of such atmospheric particles 
that need to be found at this altitude at any given
time, so that there is a chance that
at least one particle attains upward speed
around escape velocity level within the
course of one year, we find this requires as little
as one particle in
every $10^7 {\rm m}^2$ surface area 
of the atmosphere at or above this altitude.
Note that for atmospheric particles up to radius 
$r_p \stackrel{<}{\sim} 10^{-4} {\rm m}$
the estimate for the rate $R$ goes down
by only a factor two or so.
For atmospheric particles bigger than this, the rate then decreases
by a few orders of magnitude for each order of magnitude increase
in radius. This is because the particle flux of space dust 
of radius bigger than
around a millimeter decreases significantly.  

In the above analysis, the focus was on small sized space dust
in the micron to millimeter range, but the same basic
idea will extend to other sized space dust and
meteoroids.  For example there will also be on roughly
a daily basis the few
larger centimeter to meter sized meteoroids grazing past the 
Earth \citet{flynn2002,plane}.
For those grazing
above $~\approx 150 {\rm km}$,
such meteoroids could collide with
particles in the atmosphere
and send them out into space.  Its possible 
in such cases that the atmospheric
particles may even attach to the grazing
meteoriod, adding the bonus of further protection once in
the harsh space environment.
Also sizeable space dust flux has been measured 
down to nanometer and smaller sizes \citet{csfbkch}, which
could impart momentum on similar or smaller
sized particles, such as 
molecules
or tiny microbes, in the 
higher atmosphere.  

\subsection{Planetary escape of atoms and molecules comprising the higher atmosphere}

Hypervelocity space dust also has a role in the collection and exchange of
chemical constituents amongst planetary atmospheres.
This has similarities to the impact ejection 
mechanism \citet{nkmm2012}, only here for smaller scale
fast dust particles.
At altitudes above where it can abalate, $\approx 130 {\rm km}$,
an Earth-grazing space dust particle
can collide with the elements and molecules in the atmosphere,
which could stick to it and get carried
out to space.
At lower altitude, the space dust will abalate or 
fall to the Earth, releasing all its content, amongst which the
lighter elements collected from that dust particle's
journey to Earth will disperse into
the atmosphere.
To get some idea of scale,
for a space dust particle of radius $r_s$, if it travels a distance
$d$ at altitude around $z$ through the atmosphere of density
$\rho_a(z)$, it will sweep through an amount of mass 
$\Delta m = \alpha \rho_a(z) 4 \pi r_s^2 d$ in the air,
where $\alpha$ is the fraction of the atmospheric elements and
molecules that stick to the space dust particle.
The maximum mass
will be swept where $\rho_a(z)$ is largest, thus at the lowest
possible altitude safe from abalation,
so somewhere around $130-150 {\rm km}$.
Moreover once above around $150 {\rm km}$, the atmospheric
density requires an altitude increase of around $50 {\rm km}$
to decrease by an order of magnitude \citet{brekke13}. 
As such, most of the mass swept from the atmosphere by space dust will
occur in the altitude range between around $130 - 200 {\rm km}$.
Evaluating the density at
$\rho_a(150 {\rm km})$, setting
$d = 50 {\rm km}$, and $\alpha =1$ gives 
$\Delta m = \rho_a(150 {\rm km}) 4 \pi r_s^2 50 {\rm km}$.
Multiplying by the flux of space dust
with some upward velocity over the surface of the Earth
at this altitude, $f_u 4 \pi R_e^2$ and adding up the
contribution over
all sizes of space dust, we
find the mass of air
collected by space dust and escaping to space is 
around $10^{-2} {\rm g}$ every second.  This
can maybe increase by another order of magnitude by estimating
at a slightly lower altitude like $130 {\rm km}$ or extending
the distance $d$.

The thermosphere region dominantly contains oxygen, nitrogen and 
helium, so these will be the main elements that will be swept up
by space dust.  The amount escaping is much less than the
approximate $3$ kilograms of hydrogen escaping every second,
but for these heavier elements this space dust mechanism
would be competitive or even dominant to Jeans escape
and other mechanisms \citet{hd76,cz09}.
Escape via space dust will propel
these elements at high speed into space. Eventually they
could reach other terrestrial bodies, thereby
providing an exchange mechanism for atmospheric elements
amongst planets.
During the long transit time through the Solar System and beyond,
space dust will also collect the light elements, mostly
hydrogen, that are present in dilute quantities in space.
The space dust will then deliver all these light elements/molecules
to the host planetary object it eventually crashes into
at journey's end.
One possible consequence is
smaller terrestrial bodies may beable to sustain their
atmospheres longer because the incoming space dust keeps reintroducing
light elements.

\subsection{Discussion of mechanisms pushing small particles from lower to higher atmosphere}

For small, micron sized particles, there are several 
mechanisms that have been discussed
in the literature that can push them from the lower atmosphere
up to the higher attmosphere, where a collision with
fast dust conceivably could then thrust them into space.  
We will review some of these possibilities.
However first, let us explore another such mechanism which so far
has not been explored, following the same theme, that of
space dust colliding with particles at lower altitudes and pushing them
upward.  From our earlier estimates of allowed particle speeds,
we see at lower altitudes drag and heating will inhibit small
sub-millimeter sized particles from acquiring anything even
close to the escape velocity range.  Equivalently, fast space dust, as
it comes down into the lower atmosphere will also heat up and
slow down.  The heating will cause this dust to fragment.
Detailed estimates of ablation show than once the radius of the fast moving 
space dust is below around
$\approx 10^{-6} {\rm m}$, the
particle becomes efficient enough to radiate its heat
and it seizes to 
ablate \citet{plane,whipple50,hughes97,cagb10}.
If it is assumed the entire flux of space dust by the time it
reaches the lower altitudes below $100 {\rm km}$
has fragmented
into particles of radius $10^{-6} {\rm m}$, based on the flux 
data in \citet{kw86,lb93,flynn2002,plane,csfbkch}
we estimate the flux below 
this altitude would be $0.1 {\rm m}^{-2} {\rm s}^{-1}$.
This is a crude model, but it should allow us to make
some initial estimates.
For altitudes around $50 {\rm km}$ or below, atmospheric drag
forces will prohibit particles from gaining speeds beyond
tens of ${\rm m/s}$, thus attaining very little increase in
altitude and quickly succumbing to the downward force of gravity.
At or above $85 {\rm km}$ our estimates on
drag and heating show that
a single collision with space dust
could thrust a small atmospheric particle
of radius $~\approx 10^{-6} {\rm m}$ up to
$150 {\rm km}$.
This is interesting, since up to this altitude there are mechanisms
such as volcanic eruption that at least in rare cases
are known to propel
particles, and its an altitude at which small particles are observed such
as those helping to form  noctilucent clouds \citet{ludlam57}.
At $85 {\rm km}$ altitude $\sqrt{2z/R} = 0.16$, so the upward
flux of the space dust we estimate to be
$f_u \approx 2 \times 10^{-2} {\rm m}^{-2} {\rm s}^{-1}$.
If the number 
density of atmospheric particles of radius $~\approx 10^{-6} {\rm m}$
at altitude a few kilometers thickness around $85 {\rm km}$
were $~\approx 1/{\rm m}^3$, then our estimates show the space dust
collision mechanism could push enough
particles up to altitude $150 {\rm km}$ to attain the minimum
density up there to allow, based on our earlier calculation,
for the chance of a second collision to propel
at least one such particle free of the Earth's gravity in a year.

To make further progress with these estimates, a better understanding of
the distribution of small particles in the atmosphere is needed.
Up to the middle of the stratosphere, so altitudes up to 
$35 {\rm km}$, various measurements have shown there are small particles
of radius within an order of magnitude range of a micron and
in concentrations within a couple of orders of magnitude of
one per ${\rm cm}^3$ \citet{rosen64,xszzi,hht,ycjczz,ursem16}.
Measurements have even shown that amongst the particles are
bacteria \citet{ursem16,wwnr03,griffin04}.  
At the upper end of the troposphere at around $10 {\rm km}$ altitude
after hurricanes,
bacteria number concentrations were found to be 
as high as $0.1 {\rm cm}^{-3}$ \citet{dretal13}.
Even higher up at $41 {\rm km}$, bacteria have been 
detected \citet{wwnr03,ursem16}.  Further up noctilucent clouds
at altitudes of $80 - 100 {\rm km}$ provide evidence that
small particle matter must reside, although the origin
of particles this high up is argued both as terrestrial and
from space dust \citet{dld,rohatschek96,ursem16,wwnr03,ludlam57}.

There also are various mechanisms that to varying degree are known
capable of sending small particles high up into the atmosphere.  Hurricanes
and other strong weather activity can thrust particles
up the troposphere. Volcanoes can thrust ash well into 
the stratosphere \citet{wshw78,tthgr09}. 
From the powerful eruption by Krakatao in 1883
\citet{verbeek1884,sr81} there are suggestions that dust from the
volcanic ash diffused up to
$85 {\rm km}$ and has been regarded as a source for
noctilucent clouds that appeared at the time \citet{ludlam57}.
Blue jets and sprites from the tops of thunderclouds in the troposphere
have speeds in the
thousands of meters per second so offer a powerful source to thrust particles
upward into the stratosphere or higher \citet{pasko2002}.
The process of gravito-photophoresis, arising from irradiation of
particles by sunlight, has been shown can elevate micron scale
particles to altitudes upward of $80 {\rm km}$ \citet{rohatschek96}.
Mesospheric and thermospheric upward vertical winds have been
measured in the tens to hundreds
of ${\rm m/s}$ \citet{wg74,rb76,rscmla84,ketal09,ermr11}.
These mechanisms along with space dust collisions in the lower atmosphere
seem suggestive that some micron sized particles can get pushed up
well into the thermosphere.

\subsection{Shock pressure during collision}

Another concern for this mechanism is,
if atmospheric particles are thrust into space by
a hard collision, if any microbes are present in those particles
can they withstand such powerful hits.
The possibility for such atmospheric particles to 
to be destroyed through processes like fragmentation and vaporization
are not too significant up to collision speeds of interest here
$\stackrel{<}{\sim} 50 {\rm km/s}$ \citet{bd95}.
However the shock pressure created during such collisions
is a concern, with
studies showing that bacteria can survive collisions
with shock pressure at least up to $50 {\rm GPa}$ \citet{setal07,psbjac13}.  
For an initially stationary
atmospheric particle or radius $r_p$, if it is
impacted by another particle at velocity $v_s$, the
collision time would be approximately the time it takes
the impinging particle to travel the size of the
initially stationary particle, $\Delta t \sim r_p/v_s$.  
If the stationary particle emerges from
this collision with velocity approximately the same as the
incoming particle, then the force from this collision would
be $\Delta p/\Delta t = m_p v_s^2/r_p$, 
where $m_p = 4 \pi r_p^3 \rho_p/3$ is the
mass of the stationary particle and $\rho_p$ its density.
So the approximate shock pressure on the particle would
be $\sim 4 \rho_p v^2/3$.  For a particle emerging at velocity
$11.2 {\rm km/s}$ and $\rho_p = 2 \times 10^6 {\rm g/m^3}$ this leads
to a shock pressure of $\sim 3 \times 10^2 {\rm GPa}$, which is less than an
order of magnitude higher than the survival limits tested so far.
If imparting only the velocity needed to reach an orbit around the
Earth $~\approx 7.8 {\rm km/s}$ is considered, it leads to a shock pressure
$~\approx 10^2 {\rm GPa}$, which is within a factor two 
of the survival range from the tests
that have been done.
If the particle is already moving fast and just needs an
increase in velocity of a few ${\rm km/s}$ to reach escape velocity,
then the shock pressure would be well below $50 {\rm GPa}$.
Also if the collision is
with a particle containing many microbes packed together,
some may absorb the main impact of the collision, while leaving
others intact.  Moreover if the organisms undergoing the collision
are in an anhydrobiotic state, they would better withstand the
shock pressure and in addition would be greatly reduced in mass,
so also making it easier to accelerate them to high speed.
From these considerations we find that
the impact shock
is a key limiting factor in controlling the amount of
microbial life that will be able to escape the Earth from
fast space dust collisions.

\section{Discussion}

Should some microbial particles manage the perilous journey upward
and out of the Earth's gravity, the question remains how well they
will survive in the harsh environment of space. Bacterial spores
have been left on the exterior of the
International Space Station at altitude $\approx 400 {\rm km}$,
in a near vacuum environment of space, where there is nearly no
water, considerable radiation, and
with temperatures ranging from 332K on the sun side
to 252K on the shadow side,
and have survived
1.5 years \citet{hetal12}.
Other experiments with bacteria \citet{hetal01}
and lichens \citet{setal05} have shown survival
of these organisms also openly exposed to the vacuum of space,
with its radiation and extreme temperatures.
A small invertebrate animal, the tardigrade is even more
resilient, surviving extreme temperature, heat,
pressure, and radiation \citet{horikawa11} and has been
shown to survive in space \citet{jonsson08}.
Thermophiles are known to survive
in temperatures reaching up to 400K \citet{tetal08}.
Thus for microbes that manage the hypervelocity
escape from the Earth, it seems some would be hardy enough
to also survive in the region of space
nearby Earth.  If these microbes continued to journey
further out in the Solar System,
radiation levels would decrease but
temperatures would get much
more cold down to 40K at the outer part of the Solar System.
Tardigrades have been tested and shown to survive
at liquid nitrogen temperatures
77K \citet{rw92} and even near absolute zero \citet{becquerel50}.
Note that further tests of microbes
at very low temperatures and for long duration could be done at
the Large Hadron Collider facility at CERN,
which has a functioning cryogenic unit
containing liquid helium cooled down to 1.9K.

If biological constituents have been
escaping the Earth continuously, even in tiny
amounts, over its
lifespan, then
it would suggest floating out in
the Solar System there is information about
the evolutionary history of microbial life over the time of
the Earth's existence.
Once a particle escapes from Earth, it 
could circulate around the Solar System,
possibly eventually landing on another planet or even returning
back to Earth.  Such a particle would have a much better
chance for survival if upon escape, it was quickly
swept up by an oncoming comet, asteroid or other near Earth object.
Even if the escaped particle only contained
organic molecules or microbes that were killed
on their journey out of Earth, such
complex organic systems may still
act as blueprints in suitable environments to speed up
the development of life through assisting self-replication, self-organization,
abiogensis and various other potential mechanisms.

Once atoms, molecules or even bigger sized particle escape from Earth, they
could still remain within the
gravitational bound of the Solar System and circulate around,
possibly eventually landing on another planet or even returning
back to Earth.  Another possibility is the particle
gains enough speed to leave the Solar System altogether.
Since the Earth is rotating at $~\approx 30 {\rm km/s}$
around the sun, a particle
escaping the Earth generally would emerge with
speed in the tens of ${\rm km/s}$.  If the speed exceeds
around $42 {\rm km/s}$, it would be fast enough, if unencumbered and
headed in the right direction, to
escape the gravitational bound of the Solar System.
Upon emerging from the Solar System,
such a particle could still have a speed
in the same order of tens of ${\rm km/s}$,
so that over the lifespan of the Earth of
four billion years, particles emerging from Earth by
this manner in principle
could have travelled out as far as tens of kiloparsecs.
This material horizon, as could be called the
maximum distance on pure kinematic grounds that a material
particle from Earth could travel outward based on natural processes,
would cover most of our
Galactic disk, and interestingly
would be far enough out to reach
the Earth-like or potentially habitable planets that have been identified.
As such, these
estimates show the exchange
of atoms, molecules and even 
small biological constituents amongst
these distant Earth-like
planets, can not on kinematic grounds be
entirely excluded.  Numerical studies
show it would be extremely improbable for
a meteorite originating from a planet in one
solar system to fall onto a planet in another
Solar System \citet{melosh2003}.  The same conclusion is likely to
follow for the small particles propelled into
space by the space dust collision mechanism we are considering in
this paper.  Moreover the long time spent in the harsh
space environment in such distant journeys would lead to
considerable exposure to radiation and cosmic rays that
most likely would destroy any biological life.
It is possible that some of the
organic molecules contained in these small 
particles may survive. Their unique sequences 
coming from different stellar systems could therefore intermix and 
have some influence in
the development of biological processes amongst distant
solar systems. 

Collisions of huge meteorites
with the Earth are a well
known mechanism for raising large amounts of material from
the Earth out into space, some of it possibly containing
microbial life \citet{melosh85,gdlb05}.  
Although this is a potential mechanism
for throwing microbial life into space, it occurs very
rarely, on geological time scales.  This microbial life
once in the harsh space environment would have the best chances of survival if
around the time of the impact other near Earth objects,
like asteroids, comets etc..., in the Solar
System were also sweeping past the Earth and collected
the microbial debris, thus helping to protect
this life and facilitate its transfer to larger planetary bodies.
In contrast, the mechanism suggested
in this paper at best could propel only a small amount 
of Earth's biological constituents
into space.  However the influx of space dust is continuous,
so the process of expelling particles could occur
more frequently and so possibly increasing the chances
that some of it is collected
by an asteroid, comet or other near Earth object as it comes
past the Earth. 
Note that if this
space dust collision mechanism succeeded to propel just one
small particle containing life from a planetary
system like Earth into space even just every several thousands of years,
that still implies that many such events would have occurred
over geological timescales.
The key estimate made in this paper is that even for a vanishingly small
concentration of small particles harboring microbial life
in the upper atmosphere,
the space dust collision mechanism still provides
the possibility for several escape events over geological timescales.

Huge amounts
of microbial life need not have to escape from Earth, but what is
equally important is that they escape at the right time, 
so as to be collected
by a body that can allow that life to flourish and multiply.
Also the space dust collision
mechanism proposed here is as likely to be occurring 
in the present as in the past.
As such it can be tested as it occurs in the present
rather than requiring the modelling and assumptions needed
for testing huge meteor strikes from the distant past.
However for both mechanisms, whether huge meteor stike
or space dust collision, there is the common concern whether microbial life
can survive the impact of the collision. Just as many
investigations have gone in studying the meteor mechanism,
much more investigation will be needed for the space dust mechanism
proposed in this paper.
This paper has made order of magnitude estimates. 
For the application to the escape of atoms/molecules forming the atmosphere
and the escape of bigger particles, better modelling of 
Earth-grazing space dust flux would improve our estimates.
In addition, for the application to the escape of larger particles,
more information is needed about
the density of small,
micron sized and larger,
particles in the mesosphere and higher.
Direct measurement of the density
of such small particles in the higher atmosphere, into the mesosphere and
above, would benefit not just the further development
of this mechanism but would provide further knowledge
about the constituents making up the atmosphere, which may be useful
elsewhere such as climate science.
Collecting particles
by high altitude balloons,
sounding rockets, 
or higher up at $400 {\rm km}$ by
the International Space Station are possibilities.
Conceivably the vast amount of man-made space junk orbiting
around the
Earth, primarily at altitudes around $750 {\rm km}$, would contain
useful information about both the incoming flux of space dust
and outgoing flow of small particles from Earth.  
Since some of this debris has been orbiting for decades and the over
million objects are distributed all around the Earth, they contain
long time and spatially well distributed information.  Possibly
microbes have even become housed in some of it, albeit in a dormant state,
or complex organic molecules can be found there.
Remote analysis of this space junk and conceivable
retrieval of some of it in the future would allow assessing
its science content.

The influx of hypervelocity space dust
creates huge and sustained momentum flows
in the planetary atmosphere.  This paper
is the first to observe that fast space
dust particles inevitibly will sometimes collide
with particles residing
in the atmosphere and cause them to be moved around
and may thrust some
out into space.  
Hypervelocity space dust is a unique entity in planetary systems
like our Solar System,
which is able to go past and enter the
atmosphere of planets, collect samples of those planets and deposit
samples of other planets.  The entire system of fast space
dust in a planetary system thus contains
the atoms, molecules and possibly even microbial life,
from all the planets and provides a means to mix them
up amongst the different planets.  For collecting atoms and molecules
that form atmospheres, the mechanism proposed in this paper
is fairly straightforward.  For collecting life and life related
molecules this mechanism has interesting features, but
many detailed issues would still need to be studied.
The violent collisions involved in this mechanism
could make it difficult for life to remain intact.
There are several possible collision scenarios that would all need
to be explored to get a definitive answer to this problem.
But even if life itself does not remain intact,
it could still permit the complex molecules associated with
life to get propelled into space, and that is also
interesting for the panspermia process.
Since space dust is ubiquitous all over the Solar System and is
believed to exist in interstellar and probably intergalactic space,
the mechanism proposed in this paper for propelling
small particles into space could provide a universal mechanism
both for the exchange of the atomic and molecular constituents between
distant planetary atmospheres and for
initiating the first step of the 
panspermia process.

\acknowledgments
I thank Javier Martin-Torres and David Hochberg for helpful discussions.
This research was supported by the Science Technology Funding
Council (STFC).






%
%
%

\end{document}